\documentclass[a4paper,runningheads]{llncs}

\usepackage[utf8]{inputenc}
\usepackage{hyperref}
\usepackage{amsmath, amssymb}
\usepackage{xspace}
\usepackage{comment}
\usepackage{graphicx}
\usepackage{enumerate}
\usepackage[inline]{enumitem}
\usepackage{microtype}
\usepackage{subcaption}
\usepackage{todonotes}
\usepackage[capitalise]{cleveref}
\usepackage{thm-restate}
\usepackage{lineno}

\spnewtheorem{observation}{Observation}{\bfseries}{\itshape}

\spnewtheorem{myclaim}{Claim}{\bfseries}{\itshape}

\title{Outer-(ap)RAC Graphs}

\author{Henry~F{\"o}rster\inst{1}\orcidID{0000-0002-1441-4189} \and
Julia~Katheder\inst{1}\orcidID{0000-0002-7545-0730} \and
Giacomo~Ortali\inst{2}\orcidID{0000-0002-4481-698X}}

\institute{Wilhelm-Schickard-Institut f{\"u}r Informatik, Universit{\"a}t T{\"u}bingen, T{\"u}bingen, Germany \email{\{henry.foerster,julia.katheder\}@uni-tuebingen.de}
\and Department of Engineering, University of Perugia, Perugia, Italy \email{giacomo.ortali@unipg.it}}

\authorrunning{H. Förster, J. Katheder, G. Ortali}

\begin{document}
\maketitle

\begin{abstract}
    An \emph{outer-RAC drawing} of a graph is a straight-line drawing where all vertices are incident to the outer cell and all edge crossings occur at a right angle. If additionally, all crossing edges are either horizontal or vertical, we call the drawing \emph{outer-apRAC} (\emph{ap} for \emph{axis-parallel)}. 
    A graph is outer-(ap)RAC if it admits an outer-(ap)RAC drawing. We investigate the class of outer-(ap)RAC graphs. We show that the outer-RAC graphs are a proper subset of~the planar graphs with at most $2.5n-4$ edges where $n$ is the number of vertices. This density bound is tight, even for outer-apRAC graphs. Moreover, we provide an SPQR-tree based linear-time algorithm which computes an outer-RAC drawing for every given  series-parallel graph of maximum degree four. As a complementing result, we present planar graphs of maximum degree four and series-parallel graphs of maximum degree five that are not outer-RAC. Finally, for series-parallel graphs of maximum degree three we show how to compute an outer-apRAC drawing in linear time. 
    \keywords{RAC, beyond planarity, density, series-parallel graphs}
\end{abstract}

\section{Introduction}

Crossings in graph drawings are well-known to impede readability. This fact was experimentally verified by Purchase~\cite{DBLP:journals/iwc/Purchase00} in 2000. Follow-up works showed that the topology and geometry of local crossing configurations are deciding factors in how large the impact of crossings on readability actually is. Crossings~at larger crossing angles reduce readability to a lesser extent than those at smaller crossing angles~\cite{DBLP:conf/apvis/Huang07,DBLP:journals/vlc/HuangEH14,DBLP:conf/apvis/HuangHE08}. These results gave rise to the research field \emph{graph drawing beyond planarity} where graph drawings with specific requirements towards local crossing configurations have been considered. Substantial research has deepened our understanding of beyond-planar graphs; see \cite{DBLP:journals/csur/DidimoLM19,DBLP:books/sp/20/HT2020} for an overview.

In this paper, we consider \emph{right-angle-crossing drawings}, or \emph{RAC drawings} for short, which are straight-line drawings of graphs where every crossing occurs at a right angle. The RAC drawing model is directly motivated by empirical studies that gave rise to a deeper study of graph drawing beyond planarity~\cite{DBLP:conf/apvis/Huang07,DBLP:journals/vlc/HuangEH14,DBLP:conf/apvis/HuangHE08}. Hence, it comes at no surprise, that RAC drawings have been thoroughly investigated. More precisely, a first theoretical study by Didimo et al.~\cite{DBLP:journals/tcs/DidimoEL11} established a linear edge density bound shortly after Huang's initial eye tracking study~\cite{DBLP:conf/apvis/Huang07}. In addition, they showed that every graph admits a RAC drawing with $3$ bends per edge~\cite{DBLP:journals/tcs/DidimoEL11} which was shown to be the tight number of bends by Arikushi et al.~\cite{DBLP:journals/comgeo/ArikushiFKMT12}. Subsequent works on RAC drawings considered edge density~\cite{DBLP:journals/tcs/AngeliniBFK20,DBLP:journals/corr/abs-2311-06193,DBLP:conf/gd/Toth23}, area~\cite{DBLP:conf/esa/Forster020,DBLP:journals/ipl/RahmatiE20}, variants where edges are drawn as circular arcs~\cite{DBLP:conf/swat/ChaplickFK020}, simultaneous RAC drawings~\cite{DBLP:journals/jgaa/ArgyriouBKS13,DBLP:journals/jgaa/BekosDKW16} and algorithms for restricted input graphs~\cite{DBLP:conf/mfcs/AngeliniBKKP22,DBLP:journals/tcs/BekosDLMM17,DBLP:journals/comgeo/ChaplickLWZ19}. The complexity of the RAC drawing problem has first been  shown to be NP-hard~\cite{DBLP:journals/jgaa/ArgyriouBS12} and later to be $\exists \mathbb{R}$-complete~\cite{DBLP:journals/jgaa/Schaefer23a}; on the other hand, there are FPT algorithms parameterited by feedback edge number and by vertex cover number~\cite{DBLP:conf/gd/BrandGRS23}. Recently, a variant of RAC drawings called \emph{axis-parallel RAC drawings}, or \emph{apRAC drawings} for short, was introduced, in which each crossing edge has slope $\pm 1$~\cite{DBLP:conf/esa/AngeliniBK00U23}\footnote{Originally, the slopes where defined as $0$ and $\infty$ in \cite{DBLP:conf/esa/AngeliniBK00U23}. We use the rotated version $\pm 1$ which will allow us to simplify our discussion in \cref{sec:drawing}.}.

In beyond-planar graph drawing, a classical topic is to consider additional \emph{constraints} for the drawings. One of these constraints is the \emph{outer drawing} model where each vertex must be located on the outer cell of the drawing. Outer drawings may be utilized to visualize highly connected clusters in graphs~\cite{DBLP:journals/tvcg/AngoriDMPT22,DBLP:journals/tvcg/HenryFM07} which in real-world networks are often only sparsely connected to each other; see e.g.~\cite{FORTUNATO201075,doi:10.1073/pnas.122653799}. Previous research has considered outer-$k$-planar~\cite{DBLP:journals/algorithmica/AuerBBGHNR16,DBLP:journals/jgaa/Biedl22,DBLP:journals/algorithmica/HongEKLSS15,DBLP:conf/wg/HongN15}, outer-fanplanar~\cite{DBLP:journals/algorithmica/BekosCGHK17,DBLP:journals/tcs/BinucciGDMPST15} and outer-confluent graphs~\cite{DBLP:journals/jgaa/ForsterGKN21}. Surprisingly however, the existing literature only considered outer-(ap)RAC drawings with additional constraints on the placement of vertices~\cite{DBLP:journals/tcs/DehkordiEHN16,DBLP:journals/algorithmica/GiacomoDEL14} and for outer-$1$-planar graphs~\cite{DBLP:journals/ijcga/DehkordiE12}.


\paragraph{Our contribution.} We initiate the study of more general outer-(ap)RAC drawings in which vertices can be arbitrarily placed as long as they are incident to the outer cell. In the process, we prove that the outer-RAC graphs are a proper subfamily of the planar graphs in \cref{sec:topology}. Moreover, we show that certain planar graphs of low maximum degree do not admit outer-(ap)RAC drawings in \cref{sec:counter-examples}. In contrast, we provide efficient outer-(ap)RAC drawing algorithms for series-parallel graphs of low maximum degree in \cref{sec:drawing}. Finally, we conclude the paper with intriguing open questions.

\section{Preliminaries}
We assume familiarity with standard notation from graph theory, as found in~\cite{Diestel} and basic graph drawing concepts, cf.~\cite{DBLP:reference/crc/2013gd}. In this paper, we consider all graphs to be simple. Let $G=(V,E)$ be a graph. A graph is said to be \emph{cubic}, if all of its vertices have exactly degree $3$. 
In a \emph{subcubic} graph, every vertex has degree at most $3$. The terms \emph{(sub)quartic} are defined analogously for vertex degree $4$. We call a drawing $\Gamma$ \emph{planar} if in $\Gamma$ no two edges intersect except at a common endpoint. We say that a graph $G$ is \emph{planar} if it admits a planar drawing. The connected regions of the plane in a planar drawing $\Gamma$ are called \emph{faces}, the unbounded face is called \emph{outer face}. A planar drawing $\Gamma$ in which each vertex is incident to the outer face is called \emph{outerplanar}. If a graph $G$ admits an outerplanar drawing we call $G$  an \emph{outerplanar graph}. In the weak dual graph $H$ of a planar drawing $\Gamma$, each face except the outer face is represented by a vertex and faces $f_1, f_2$ are connected in $H$ if and only if $f_1$ and $f_2$ share an edge in $\Gamma$. For an embedded outerplanar graph, the weak dual graph is a forest.

Similarly, the connected regions of the plane in a \emph{non-planar} drawing $\Gamma$ are called \emph{cells} and the unbounded cell is called \emph{outer cell}. Consider a straight-line drawing $\Gamma$, i.e., each edge is represented by a single segment. If in $\Gamma$, all crossings occur at a right angle and all vertices are located on the outer cell, we call $\Gamma$ \emph{outer-RAC}. If additionally, all crossing edges have slope $\pm 1$, we call $\Gamma$ \emph{outer-apRAC}. Moreover, we call a graph \emph{outer-(ap)RAC} if it admits an outer-apRAC drawing.

An \emph{SPQR-tree} $\mathcal{T}$ of a graph $G$ describes a uniquely defined decomposition of $G$ according to its separation pairs~\cite{DBLP:journals/algorithmica/BattistaT96,DBLP:journals/siamcomp/BattistaT96}, which can be computed in linear time~\cite{DBLP:conf/gd/GutwengerM00}. A node $\mu$ in $\mathcal{T}$ is associated with a graph $skel(\mu)$ called the \emph{skeleton} of $\mu$ which consists of \emph{virtual edges}  between the vertices of separation pairs in $G$ and at most one edge of $G$. Each virtual edge corresponds to at least one path between its endpoints in $G$. The pair of vertices $s_\mu, t_\mu$ separating the component represented by $\mu$ in $G$ are called the \emph{poles} of $\mu$. The vertices $s_\mu, t_\mu$ are connected by the \emph{parent virtual edge}, which corresponds to a virtual edge in the parent of $\mu$ in $\mathcal{T}$. Based on the structure of its skeleton, a node $\mu \in \mathcal{T}$ is of one of four types:
\begin{itemize}
    \item \emph{S-node}: $skel(\mu)$ forms a cycle of at least three virtual edges, including the parent virtual edge between $s_\mu$ and $t_\mu$
    \item \emph{P-node}: $skel(\mu)$ is comprised of at least three parallel virtual edges between $s_\mu$ and $t_\mu$ one of which is the parent virtual edge
    \item \emph{Q-node}: $skel(\mu)$ contains the parent virtual edge and another edge between $s_\mu$ and $t_\mu$, representing an actual edge in $G$. If $\mu$ is the root of $\mathcal{T}$, the virtual edge of the skeleton instead corresponds to its unique child node $\nu$.
    \item \emph{R-node}: $skel(\mu)$ is triconnected and contains the poles $s_\mu$ and $t_\mu$. All its edges are virtual, the virtual edge between $s_\mu$ and $t_\mu$ is its parent virtual edge.
\end{itemize}
The SPQR-tree $\mathcal{T}$ of a graph $G$ is by definition rooted at a $Q$-node. Moreover, all its leaves are $Q$-nodes. Also observe that two $S$-nodes are never connected to each other and the same holds for two $P$-nodes in $\mathcal{T}$. The subtree rooted at a node $\mu$ induces the so-called \emph{pertinent graph} $pert(\mu)$, which is a subgraph in $G$ obtained from merging the parent virtual edge in the skeleton of each node in the subtree of $\mathcal{T}$ rooted at $\mu$ with the corresponding virtual edge in the skeleton of its respective parent node.  
A \emph{series-parallel graph}, or short \emph{SP-graph}, is a biconnected graph\footnote{In the literature there exists another recursive definition for series-parallel graphs~\cite{DBLP:journals/iandc/Eppstein92}. However, one can obtain biconnectivity by a  parallel composition with a single edge.} whose SPQR-tree contains no $R$-nodes. Since the skeleton of both P- and S-nodes are always planar, SP-graphs are always planar. 

\begin{lemma}
    \label{lem:rerooting}
    Let $G$ be an SP-graph with $n \geq 3$ vertices and let $\mathcal{T}$ be its SQPR-tree rooted at any $Q$-node. Then, $\mathcal{T}$ contains an $S$-node $\mu$  such that all children of $\mu$ are $Q$-nodes.
\end{lemma}

\begin{proof}
    If $G$ contains no $P$-node the statement follows immediately. Otherwise, since all leaves of $\mathcal{T}$ are $Q$-nodes, we find a $P$-node $\mu_p$ by traversing $\mathcal{T}$ top-down, such that the subtree of $\mathcal{T}$ rooted at $\mu_p$ contains no other $P$-node. By simplicity, $\mu_p$ has at most one $Q$-node child and hence at least one $S$-node child $\mu_s$ whose children are all $Q$-nodes. 
\qed\end{proof}

We will make use of \cref{lem:rerooting} to simplify the discussion of our algorithmic results in \cref{sec:drawing} as follows. Let $\mu_s$ be an $S$-node such that all its children are $Q$-nodes in $\mathcal{T}$. We can now root $\mathcal{T}$ at one of the $Q$-node children of $\mu_s$, denoted as $\mu_r$. After rerooting, we have that $\mu_s$ is the unique child of $\mu_r$. 

\begin{corollary} \label{prop:subcubic-root}
    Let $G$ be an SP-graph and $\mathcal{T}$ be its SPQR-tree.  $\mathcal{T}$  can be rooted at a $Q$-node $\mu_r$ with unique child $\mu_s$ such that $\mu_s$ is an $S$-node and at most one child $\mu_p$ of $\mu_s$ is a $P$-node.
\end{corollary}

\section{Topological Results}
 \label{sec:topology}
\begin{figure}[t]
    \centering
    \begin{subfigure}{0.4\textwidth}
        \centering
        \includegraphics[width=\linewidth]{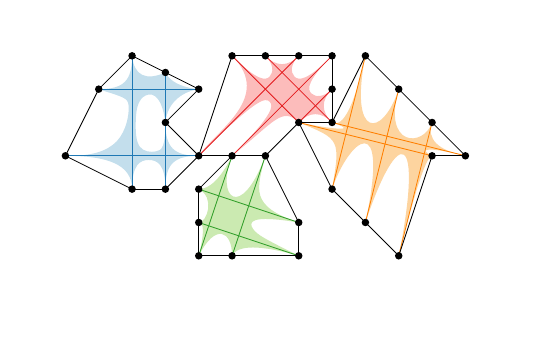}
        \subcaption{}
        \label{fig:blocks-a}
    \end{subfigure}
    \hfil
    \begin{subfigure}{0.4\textwidth}
        \centering
        \includegraphics[page=4,width=\linewidth]{outer-rac-topology.pdf}
        \subcaption{}
        \label{fig:blocks-b}
    \end{subfigure}
    \caption{(a) An outer RAC graph with its crossing edges being decomposed into blocks (colored edges) and the outlines of the blocks (colored regions). (b) Illustration for the proof of~\cref{thm:planar}.}
    \label{fig:blocks}
\end{figure}

In this section, we provide some topological results. A useful tool for these results will be the notion of \emph{blocks} of crossing edges in a given outer-RAC drawing $\Gamma$ of a graph $G=(V,E)$. To this end, consider the \emph{crossing graph} $C(\Gamma)$ of $\Gamma$ which contains a vertex for each edge in $E$ and an edge $(e_1,e_2) \in E \times E$ if and only if $e_1$ and $e_2$ cross in $\Gamma$. A \emph{block} is any maximal connected vertex set in the transitive closure of $C(\Gamma)$; see also \cref{fig:blocks}. In particular, since in RAC drawings only edges drawn with perpendicular slopes cross, each block contains edges of only two perpendicular slopes. Thus, we can partition the edges of block $B$ obtaining \emph{slope sets} $B_1, B_2 \subset B$ with $B_1 \cap B_2 = \emptyset$ such that each pair of edges of $B_i$ for $i \in {1, 2}$ does not cross. Moreover, since we consider the outer-RAC setting, all endpoints of edges of the same block are located on the outer face. Since in addition, edges assigned to different blocks do not cross  by definition, we can cover the interior of a RAC drawing with regions called \emph{outlines} of blocks that contain all crossing edges as follows. Consider a block $B \subset E$. Sort the endpoints of the edges in $B$ according to their occurrence in a clockwise cyclic walk along the outer face $f_\circ$ of $\Gamma$ and enumerate them by $v_1,\ldots,v_k$. Then, there necessarily exists a closed cycle $O(B)$, called the \emph{outline} of $B$, such that \begin{enumerate*}
    \item $O(B)$ is disjoint from the interior of $f_\circ$ (in other words, each point $O(B)$ is either inside a bounded cell of $\Gamma$ or on the boundary of $f_\circ$),
    \item $O(B)$ does not cross any edges except at their endpoints,
    \item $O(B)$ contains all edges of $B$ but no other edge of $\Gamma$, and
    \item $O(B)$ traverses $v_1,\ldots,v_k$ in order.
\end{enumerate*}
More precisely, when traversing along $O(B)$ from $v_i$ to $v_{i+1}$, one can follow the clockwise last\footnote{When considering the edges incident to $v_i$ in clockwise order starting from the direction in which we reach $v_i$ along $O(B)$ coming from $v_{i-1}$.} edge of $B$ incident to $v_i$ up to its first crossing and then following the edge encountered up to its next crossing where we again follow the crossing edge; see also \cref{fig:blocks-a}. Necessarily, due to the fact that the drawing is outer-RAC, a repeated application of this method will finish at $v_{i+1}$.  

\begin{theorem}
    \label{thm:planar}
    Let $G$ be an outer-RAC graph. Then $G$ is planar. 
\end{theorem}

\begin{proof}
Consider any outer-RAC drawing $\Gamma$ of $G$. We will describe how to obtain a planar drawing of $G$. First, extend $\Gamma$ by all the outlines of blocks obtaining a drawing $\Gamma'$. By construction, $\Gamma'$ is still bounded by the outer cycle $C$ which contains all vertices. Moreover, the outlines of blocks do not cross each other by construction. We now copy the interior of $C$ to the outside. Observe that in this copying operation we disregard the geometry but maintain the topology of the drawing. We now have two copies of each edge, each staying within its respective block outline either in the interior or the exterior of $C$. For each block $B$, consider now the slope sets $B_1$ and $B_2$. We will remove the copies of $B_1$ in the interior of $C$ and the copies of $B_2$ in the exterior of $C$. Since edges of the same slope set do not cross, we obtain a planar drawing of $G$ by removing the block outlines; see~\cref{fig:blocks-b} for an illustration.
\qed\end{proof}

One may wonder if in fact every planar graph is also outer-RAC. We will show that this is not the case by providing a density upper bound for outer-RAC graphs. As an intermediate step, we consider a special subclass of outer-RAC graphs. We call a graph \emph{bounded block graph} if it can be partitioned into a Hamiltonian cycle $H$ and a set of edges $E_c$ and  admits a \emph{bounded block drawing} $\Gamma$, that is, a drawing where \begin{enumerate*}
    \item each edge of $E_c$ is drawn straight-line and crosses only at right angles,
    \item $H$ forms the crossing-free outer boundary of $\Gamma$ and is not necessarily drawn straight-line, and
    \item all edges of $E_c$ belong to a single block.
\end{enumerate*} 

\begin{lemma}\label{lem:boundedblock}
    Let $G$ be a bounded block graph with $n$ vertices. Then, $G$ has at most $2n-2$ edges.
\end{lemma}

\begin{proof}
    Let $\Gamma$ be a bounded block drawing of $G$.
    The crossing edges of $G$ form a block $B$ that can be partitioned into two slope sets $B_1$ and $B_2$. In addition, $G$ contains only the plane cycle $C$ which is topologically equivalent to the outline $O(B)$. We now consider for $i \in {1, 2}$ the subgraph $B'_i=B[C \cup B_i]$ of $G$ induced by the edges of $C$ and $B_i$. Recall, that all edges of $B_i$ are parallel, i.e., $B_i'$ is an embedded outerplanar subgraph of $\Gamma$. We simplify $B_i'$ obtaining $B_i^*$ by replacing each maximal path $p$ of edges of $C$ on the boundary of the same internal face by a single edge unless this creates a parallel edge with an edge of $B_i$ in which case we replace $p$ by a path of length $2$. Hence, each internal face of $B_i^*$ contains exactly $2$ edges not belonging to $B_i$. We claim that for each internal face $f$ of length $4+k$, we can assign $k$ triangular faces. To this end, recall that the weak dual graph $T$ that has a vertex for each internal face and an edge between faces sharing an edge  is in fact a forest.  Since each edge of $B_i$ necessarily is part of two faces, $f$ has degree $2+k$ in $T$. Moreover, each triangular face has only one edge of $B_i$, i.e., degree $1$ in $T$. Thus, such an assignment is possible in such a way that two triangular faces $t_1$, $t_2$ remain unassigned. We now root $T$ at $t_1$ and count the number of vertices and edges in an in-order processing of $T$.

    At the root, we encounter $3$ vertices and $3$ edges. If we encounter another face $f$, it shares $2$ vertices and $1$ edge with its parent. That is, if the face has length $k$ with $k \geq 3$, we find $k-2$ additional vertices and $k-1$ additional edges. Thus, writing $f_j$ for the number of faces of length $j$, we obtain  the number of vertices $n_i^*$  and edges $m_i^*$ of $B_i^*$:
    \begin{align}
        \label{eq:mstar}
        m_i^*& = 3 + 2 + \sum_{j=4}^\infty \big( (j-1)\cdot f_j + f_j\cdot2(j-4) \big)= 5 + \sum_{j=4}^\infty (3j-9)f_j\\
        \label{eq:nstar}
        n_i^*& = 3 + 1 + \sum_{j=4}^\infty \big((j-2)\cdot f_j + f_j\cdot(j-4)\big)=4+ \sum_{j=4}^\infty(2j-6)f_j      
    \end{align}
    The additive constants refer to faces $t_1$ and $t_2$ whereas in the sum expressions we add both the vertices of faces of length $j$ as well as the assigned triangular faces. 
    
    We can now transfer back to an accounting for graph $B_i'$ by subdividing the edges replacing paths of non-crossed edges suitably. Each such subdivision creates another edge and another vertex. Writing $s$ for the number of such subdivisions and $m_i'$ and $n_i'$ for the number of vertices and 
    edges of $B_i'$, respectively, we   can refine \eqref{eq:mstar} and \eqref{eq:nstar} as follows:
    \begin{align}
    n_i'&= n_i^* + s = 4 + s + \sum_{j=4}^\infty(2j-6)f_j \Leftrightarrow \sum_{j=4}^\infty(2j-6)f_j=n_j'-4-s
    \\
    \nonumber m_i'&= m_i^* + s = 5 + s + \sum_{j=4}^\infty(3j-9)f_j =  5 + s + \frac{3}{2}\sum_{j=4}^\infty(2j-6)f_j\\& \label{eq:mprime}= 5 + s + \frac{3}{2}(n_i' -4 - s) \leq \frac{3}{2} n_i' - 1 
    \end{align}

    Now we compute the number of edges $m$ of $G$. Necessarily, we have $n=n_1'=n_2'$. Moreover, both $B_1'$ and $B_2'$ contain the outer cycle on $n$ vertices. Thus,
    \begin{equation}
        \label{eq:mb}
        m=m_1'+m_2' - n \leq \frac{3}{2} n -1 + \frac{3}{2} n - 1 - n = 2n - 2.
    \end{equation}
\end{proof}

\begin{restatable}{theorem}{densityshit}\label{thm:density-upperbound}
    Let $G$ be an outer-RAC graph with $n$ vertices. Then, it has at most $m \leq 2.5n-4$ edges. Moreover, if $G$ has exactly $2.5n-4$ edges, it can be decomposed into bounded blocks $B_1,\ldots,B_k$ such that for $i \in \{1,\ldots,k\}$
    \begin{enumerate*}
        \item $B_i$ shares exactly one edge with the subgraph induced by $B_1,\ldots,B_{i-1}$,
        \item $B_i$ is isomorphic to $K_4$. 
    \end{enumerate*}
\end{restatable}

\begin{proof}[Sketch of Proof]
    We investigate any outer-RAC drawing $\Gamma$ of $G$. We first augment $G$ by inserting all outlines of blocks defined by $\Gamma$ and by triangulating the faces bounded only by crossing-free edges. We obtain a supergraph $G'$ with an associated drawing $\Gamma'$, in which all crossing edges are straight-line. 
    Then, we observe that $G'$ can be obtained starting from a bounded block graph $G_0'$ by an iterative procedure. In each step, we have a graph $G_{i-1}'$ and obtain $G_{i}'$ by merging $G_{i-1}'$ with a new bounded block graph $B_{i}$ either at a single vertex or at an edge. 
    Finally, we show inductively that in this procedure the upper bound on the number of edges holds. See \cref{app:densityshit} for details.
\qed\end{proof}

\cref{thm:density-upperbound} already also describes a potential matching density lower bound. In fact, the existing literature already establishes that it is indeed outer-RAC. Namely, Dehkordi and Eades proved that every outer-$1$-planar graph is also outer-RAC~\cite{DBLP:journals/ijcga/DehkordiE12} whereas Didimo~\cite{DBLP:journals/ipl/Didimo13} and Auer et al.~\cite{DBLP:journals/algorithmica/AuerBBGHNR16} independently found a lower bound of $2.5n-4$ edges for outer-$1$-planar graphs. Here we strengthen their result by explicitly noting that it generalizes to outer-apRAC. 

\begin{theorem}[\cite{DBLP:journals/algorithmica/AuerBBGHNR16,DBLP:journals/ijcga/DehkordiE12,DBLP:journals/ipl/Didimo13}]\label{thm:density-lowerbound}
    There is an infinitely large family of outer-apRAC graphs with $n$ vertices and $2.5n-4$ edges.
\end{theorem}

\begin{figure}[t]
    \centering
    \begin{subfigure}{.32\textwidth}
        \centering
        \includegraphics[page=2]{outer-rac-topology.pdf}
        \subcaption{}
        \label{fig:lowerbound}
    \end{subfigure}
    \hfil
    \begin{subfigure}{.32\textwidth}
        \centering
        \includegraphics[page=1]{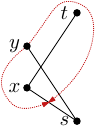}
        \subcaption{}
        \label{fig:twopaths}
    \end{subfigure}
    \hfil
    \begin{subfigure}{.32\textwidth}
        \centering
        \includegraphics[page=2]{outer-rac-obstructions.pdf}
        \subcaption{}
        \label{fig:threepaths}
    \end{subfigure}
    \caption{Illustrations for the proof of (a)~\cref{thm:density-lowerbound} (b)~\cref{lem:two-parallel-short-paths} and (c)~\cref{lem:three-parallel-short-paths}.}
    
\end{figure}

\begin{proof}
    $K_4$ is clearly outer-apRAC, e.g., place the four vertices at coordinates $(0,0)$, $(0,1)$, $(1,0)$ and $(1,1)$. Since the outer cycle of this drawing is square-shaped, we can form a chain of such $K_4$'s by identifying the left edge of a copy with the right edge of another one; see \cref{fig:lowerbound}. In this drawing, only edges of slopes $1$ and $-1$ cross and all vertices are on the outer face, i.e., it is outer-apRAC.
\qed\end{proof}

Also note that the edge density bound differs from a tight bound of $2n-2$ for \emph{circular RAC drawings}, in which all vertices are constrained to lie on a circle~\cite{DBLP:journals/tcs/DehkordiEHN16}.

\section{Obstructions for Outer-RAC Graphs}
\label{sec:counter-examples}
\cref{thm:density-upperbound} already provides examples of planar graphs that admit no outer-RAC drawings. In this section, we provide planar graphs that admit no outer-RAC drawing despite being of low enough density. To this end, we first observe that parallel short paths admit only a few specific topologies in outer-RAC drawings.

\begin{lemma}\label{lem:two-parallel-short-paths}
    Let $G$ be a graph consisting of two paths $p_1=(s,x,t)$ and $p_2=(s,y,t)$ with $x \neq y$. Then, in any outer-RAC drawing $\Gamma$, $p_1$ and $p_2$ are not self-intersecting and one of the following holds:
    \begin{enumerate}[label={P2.\arabic*},left=\labelsep]
        \item\label{case:shortpaths:1} $s$, $x$, $t$ and $y$ occur in this order along the outer cycle of $\Gamma$. Moreover, $p_1$ and $p_2$ do not intersect.
        \item\label{case:shortpaths:2} $s$, $x$, $y$ and $t$ occur in this order along the outer cycle of $\Gamma$. Moreover, $p_1$ and $p_2$ intersect exactly once at edges $(s,y)$ and $(t,x)$.
        \item\label{case:shortpaths:3} $s$, $t$, $x$ and $y$ occur in this order along the outer cycle of $\Gamma$. Moreover, $p_1$ and $p_2$ intersect  exactly once at edges $(s,x)$ and $(t,y)$.
    \end{enumerate}
\end{lemma}

\begin{proof}
First, observe that each of $p_1$ and $p_2$ cannot cross itself as  $p_1$ and $p_2$ each contain only two straight-line edges that share an endpoint.

In the following, assume that we already have an outer-RAC drawing of $p_1$. We investigate how $p_2$ can be added while maintaining that the drawing is outer-RAC. First, if $p_1$ and $p_2$ do not cross, we arrive at the configuration described in Case~\ref{case:shortpaths:1}. Second, assume that $p_1$ and $p_2$ cross. Since the pairs of edges $(s,x),(s,y)$ and $(t,x),(t,y)$ share a common endpoint, they cannot cross each other. Hence, only edge pairs $(s,x),(t,y)$ and $(s,y),(t,x)$ may cross. If exactly one of these edge pairs cross, we arrive at one of Cases~\ref{case:shortpaths:2} and~\ref{case:shortpaths:3}.

Hence, it remains to consider the case where  $(s,x)$ and $(t,y)$ as well as $(s,y)$ and $(t,x)$ cross. Assume that this is possible. Since $(s,x)$ and $(t,y)$ cross, we have that $s$, $x$, $y$ and $t$ either occur in this order around the outer cycle of $\Gamma$ or in the reversed order $s$, $t$, $x$ and $y$. We assume w.l.o.g. that the order is $s$, $x$, $y$ and $t$. Now, observe that $(x,t)$ and $y$ are separated by at least one vertex in both orientations of the outer cycle of $\Gamma$. Since $(s,y)$ must cross $(x,t)$, adding it removes at least one vertex from the outer cycle (see \cref{fig:twopaths}); a contradiction. 
\qed\end{proof}

For three such short paths, we obtain yet a different result:

\begin{lemma}\label{lem:three-parallel-short-paths}
    Let $G$ be a graph consisting of three paths $p_1=(s,x,t)$, $p_2=(s,y,t)$ and $p_3=(s,z,t)$ with $x \neq y$, $y \neq z$ and $x \neq z$. Then, in any outer-RAC drawing $\Gamma$, $p_1$, $p_2$ and $p_3$ are not self-intersecting, two paths, say $p_1$ and $p_2$, cross, whereas $p_3$ is crossing-free, and one of the following holds:
    \begin{enumerate}[label={P3.\arabic*},left=\labelsep]
        \item\label{case:shortpaths:4} $s$, $x$, $y$, $t$ and $z$ occur in this order along the outer cycle of $\Gamma$. Moreover, $p_1$ and $p_2$ intersect exactly once at edges $(s,y)$ and $(t,x)$.
        \item\label{case:shortpaths:5} $s$, $z$, $t$, $x$ and $y$ occur in this order along the outer cycle of $\Gamma$. Moreover, $p_1$ and $p_2$ intersect  exactly once at edges $(s,x)$ and $(t,y)$.
    \end{enumerate}
    Moreover, if $G$ is subgraph of an outer-RAC graph $G'$, $p_1$ and $p_2$ are not crossed by any edge not belonging to $G$.
\end{lemma}

\begin{proof}
    Consider any outer-RAC drawing $\Gamma$ of $G$.
    Using \cref{lem:two-parallel-short-paths}, we know that each pair of paths can only be realized in three different ways. First note that necessarily two paths must cross as otherwise $p_1$ and $p_2$ form a cycle that w.l.o.g. contains $p_3$; i.e., $z$ is not on the outer cycle.

    Thus, w.l.o.g. $p_1$ and $p_2$ cross according to Case~\ref{case:shortpaths:2} of \cref{lem:two-parallel-short-paths} (Case~\ref{case:shortpaths:3} is symmetric) and let $c$ denote the point where $(s,y)$ and $(t,x)$ cross. Assume for a contradiction that any edge $e^*$ not belonging to $p_1$ or $p_2$ crosses $p \in \{p_1,p_2$\}. Note that $e^*$ may be part of $p_3$ or of a supergraph $G'$ of $G$. First, since $\Gamma$ is RAC, $e^*$ cannot cross $p$ at $c$. Next, observe that $p_1$ and $p_2$ induce two triangular regions $T_1= \triangle sxc$ and $T_2=\triangle cyt$ which have a right angle at $c$; see \cref{fig:threepaths}. Moreover, all edges  of $p_1$ and $p_2$ are entirely on the boundary of $T_1$ and $T_2$. That is, if $e^*$ intersects an edge $e_p$ of $p_1$ or $p_2$, it is partially located inside $T \in \{T_1,T_2\}$. In fact, the edge $e^*$ that intersects $e_p$ must cross $T$ twice as none of its endpoints can be located in $T$ as then it would not be on the outer face. However, because $T$ is triangular, it contains no two parallel bounding segments; a contradiction.

    Since $p_3$ crosses neither $p_1$ nor $p_2$, Case~\ref{case:shortpaths:4} and Case~\ref{case:shortpaths:5} arise if $p_1$ and $p_2$ cross according to Case~\ref{case:shortpaths:2} and Case~\ref{case:shortpaths:3}, respectively.
\qed\end{proof}


\begin{theorem}\label{thm:counter-examples:1}
    There is a SP-graph of maximum degree $5$ that is not outer-RAC.
\end{theorem}

\begin{proof}
    Consider the graph $K_{2,5}$. It is a parallel composition of five paths of length $2$, i.e., it is series-parallel. By \cref{lem:three-parallel-short-paths}, it follows directly that it cannot admit an outer-RAC drawing.
    \qed \end{proof}

\begin{theorem}\label{thm:counter-examples:2}
    There is a planar triconnected graph of maximum degree $4$ that is not outer-RAC. 
\end{theorem}
\begin{proof}
    Consider the octahedral graph $G$. It consists of a $K_{2,4}$ composed of four parallel short paths  $p_1=(s,x_1,t)$, $p_2=(s,x_2,t)$, $p_3=(s,x_3,t)$ and $p_4=(s,x_4,t)$ and a cycle on $x_1,x_2,x_3,x_4$. By \cref{lem:three-parallel-short-paths}, in any of the outer-RAC drawings $\Gamma$ of $G$, w.l.o.g. we have  that its vertices occur in the order $s,x_1,x_2,t,x_3,x_4$ along the outer cycle of $\Gamma$. Since there is a cycle on $x_1,x_2,x_3,x_4$ present in $G$, $x_1$ is adjacent to $x^* \in \{x_3,x_4\}$. Clearly, edge $(x_1,x^*)$ must cross the drawing of the $K_{2,4}$ as otherwise $s$ or $t$ cannot be on the outer cycle. But then, $(x_1,x^*)$ crosses $(x_2,s)$ which, by \cref{lem:three-parallel-short-paths}, is already crossed by edge $(x_1,t)$; a contradiction.  
\qed\end{proof}

    In particular, \cref{thm:counter-examples:1,thm:counter-examples:2} motivate us  to study SP-graphs of maximum degree four. Namely, SP-graphs are exactly the planar graphs that contain no subdivision of triconnected graphs, whereas our counterexamples for SP-graphs are of maximum degree five. In  \cref{sec:drawing}, we will provide drawing algorithms for such graphs that draw the graph according to its SPQR-tree in a top-down fashion. In particular, for S-nodes, we will realize the skeleton without crossings. In \cref{app:necessarycrossingshit}, we show that this may not be guaranteed for maximum degree four SP-graphs if we restrict the drawings to be outer-apRAC.



\section{Outer-RAC Drawings for Bounded Degree SP-Graphs}
\label{sec:drawing}

\begin{theorem} \label{thm:sp-degree-3}
    Let $G$ be a biconnected SP-graph with maximum degree $3$. An outer-apRAC drawing of $G$ can be computed in $\mathcal{O}(n)$ time.
\end{theorem}
\begin{proof}
Let $G$ be a subcubic biconnected SP-graph and $\mathcal{T}$ be its SPQR-tree, hence $\mathcal{T}$ contains no $R$-nodes.
In order to construct the outer-apRAC drawing $\Gamma$ of $G$, we perform a top-down pre-order visit of $\mathcal{T}$ and draw the skeleton of each visited node  depending on its type. Let $\mu$ be the vertex currently processed. We draw the virtual edges of $skel(\mu)$ corresponding to child nodes $\nu_1,\dots,\nu_k$ of $\mu$. 
In this process, we also place the poles of its children, $s_{\nu_i}$ and $t_{\nu_i}$, $1 \leq i \leq k$. Also, when we draw the parent virtual edge corresponding to a node $\mu$, we define a \emph{reserved region} $\mathcal{R}(\mu)$, in which we will draw the skeleton of $\mu$ (aside from $s_\mu$ and $t_\mu$) later. More precisely, $\mathcal{R}(\mu)$ is defined as the intersection of three half planes, one being the closed half plane left of a line through the poles $s_\mu$ and $t_\mu$ of $\mu$ with $s_\mu$ lying above $t_\mu$, the second being the open half plane  below a horizontal line through $s_\mu$ and the third being the open half plane above a horizontal line through $t_\mu$; see~\cref{fig:region-init-1} for an illustration. 
Further, we maintain the following invariants:
\begin{enumerate}[label={I.\arabic*},left=\labelsep]
    \item \label{inv:sc-ve-vertical} Virtual edges in $skel(\mu)$ of an already processed node $\mu$ that correspond to a not yet processed child node of $\mu$ are drawn vertically.
    \item \label{inv:sc-res-region} Let $\mu$ be a not yet processed non-$Q$-node whose parent node in $\mathcal{T}$ has been processed. Then $\mathcal{R}(\mu)$ is free, i.e., it contains only 
    the  edge $(s_\mu,t_\mu)$.
    \item \label{inv:sc-45deg} 
    All crossing edges cross at right angles and have either slope $1$ or $-1$.
    \item \label{inv:sc-outer} Every already drawn vertex is incident to the outer cell. 
\end{enumerate}

\begin{figure}[t]
    \begin{subfigure}[b]{.2\textwidth}
    \centering    
    \includegraphics[width=\linewidth,page=1]{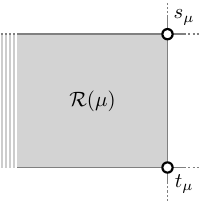}
    \subcaption{}
    \label{fig:region-init-1}
    \end{subfigure}
    \hfil
    \begin{subfigure}[b]{.18\textwidth}
    \centering
    \includegraphics[width=\linewidth,page=2]{outer-rac-sp}
    \subcaption{}
    \label{fig:region-init-2}
    \end{subfigure}
    \hfil
    \begin{subfigure}[b]{.25\textwidth}
    \centering
    \includegraphics[width=\linewidth,page=3]{outer-rac-sp}
    \subcaption{}
    \label{fig:region-init-3}
    \end{subfigure}  
    \caption{(a)~Invariants~\ref{inv:sc-ve-vertical} and~\ref{inv:sc-res-region}. (b) and (c)~initialization of \cref{thm:sp-degree-3}.}
    \label{fig:region-init}
\end{figure}
%

\smallskip 

We describe   how to compute an initial partial drawing adhering to \ref{inv:sc-ve-vertical}--\ref{inv:sc-outer} 

\paragraph{Initialization.}  Observe that if $\mathcal{T}$ contains no $P$-node, $G$ is a cycle and therefore outerplanar and thus also outer-apRAC. Hence in the following, we can assume that $G$ contains at least one $P$-node. We then root $\mathcal{T}$ according to~\cref{prop:subcubic-root}; see ~\cref{fig:region-init-2}. Now, the root $\mu_r$ has as a child an $S$-node $\mu_s$ which in turn has exactly one $P$-node child $\mu_p$. In $G$, according to the skeleton of $\mu_s$, we have that $\mu_r$ and the $Q$-node children of $\mu_s$ form a path $P$ connecting the poles of $\mu_p$. As $P$ contains at least two edges ($\mu_r$ and at least one $Q$-node child of $\mu_s$) we draw the path $P$ such that the poles of $\mu_p$ are vertically aligned as depicted in ~\cref{fig:region-init-3}, maintaining \ref{inv:sc-ve-vertical} for the virtual edge corresponding to $\mu_p$. As there are no crossings and $P$ is drawn outside of $\mathcal{R}(\mu_p)$, \ref{inv:sc-res-region},~\ref{inv:sc-45deg} and~\ref{inv:sc-outer} are guaranteed. 

\smallskip

 \noindent Next, we show how to handle a non-root node $\mu$ in the top-down traversal of  $\cal T$.

\paragraph{$\mu$ is a $Q$-node.}   
    $skel(\mu)$ consists of the poles $s_\mu$ and $t_\mu$ and an edge $e=(s_\mu,t_\mu)$ which is an edge in $G$. We draw $e$ as a vertical line connecting its endpoints, guaranteeing \ref{inv:sc-outer}. As there are no crossing edges and as $\mu$ is a leaf in $\mathcal{T}$, \ref{inv:sc-res-region} and~\ref{inv:sc-45deg} are also maintained. Since $\mu$ contains no further virtual edge, \ref{inv:sc-ve-vertical} is ensured.  
    
\paragraph{$\mu$ is a $P$-node.} Since $G$ is subcubic, $\mu$ has exactly two child nodes in $\mathcal{T}$, where each is either of type $S$ or type $Q$. Moreover, according to \ref{inv:sc-ve-vertical}, $(s_\mu,t_\mu)$ is a vertical segment and $\mathcal{R}(\mu)$ is free according to \ref{inv:sc-res-region}.
While processing $\mu$, we will partially draw the pertinent graph of each $S$-node child and remove the drawn edges from its skeleton. The remaining edges of the respective $S$-node child are then drawn in the recursive case. 
First, assume that $\mu$ has a single $S$-node child $\nu_1$ and a $Q$-node child $\nu_2$. Then the virtual edges $e_1 = (s_\mu,s'_\mu) \in skel(\nu_1)$ and $e_2 = (t_\mu,t'_\mu) \in skel(\nu_1)$  are $Q$-nodes, due to the maximum vertex degree. We delete $e_1$ and $e_2$ from $skel(\nu_1)$ and reassign the poles $s_{\nu_1}$ to $s_\mu'$ and $t_{\nu_1}$ to $t_\mu'$. If the modified $skel(\nu_1)$ is empty, we simply draw $e_1$ and $e_2$ inside the free region $\mathcal{R}(\mu)$. Otherwise, the edges $e_1$ and $e_2$ are then drawn as depicted in~\cref{fig:subcubic-p-1} within $\mathcal{R}(\mu)$. Further, we add a virtual edge $e = (s'_\mu,t'_\mu)$ to $skel(\mu)$ which represents the modified node $\nu_1$. The reserved region of $\nu_1$ is defined as $\mathcal{R}(\nu_1)$ and free as it is a subset of $\mathcal{R}(\mu)$. For $\nu_2$, the reference edge is $(s_\mu,t_\mu)$ which  is already drawn vertically. Hence, we maintain \ref{inv:sc-ve-vertical},~\ref{inv:sc-res-region},~\ref{inv:sc-45deg} and ~\ref{inv:sc-outer}.

Second, consider the case that $\mu$ has two $S$-node children $\nu_1$ and $\nu_2$. By the maximum vertex degree, the virtual edges in $skel(\nu_1)$ and $skel(\nu_2)$ incident to $s_\mu$ and $t_\mu$ correspond to $Q$-node children $\xi_1, \xi_2$ of $\nu_1$ and $\xi_3, \xi_4$ of $\nu_2$. 
Further, let $e_i = (u_i,v_i)$ be the virtual edge in $skel(\nu_i)$ corresponding to $\xi_i$, such that $s_\mu = u_1 = u_3$ and $t_\mu = u_2 = u_4$. We remove the edges $e_1,e_2$ from $skel(\nu_1)$ and redefine the poles $s_{\nu_1}$ as $v_1$ and $t_{\nu_1}$ as $v_2$. Similarly, we remove $e_3,e_4$ from $skel(\nu_2)$ such that $s_{\nu_2}= v_3$ and $t_{\nu_2}=v_4$. Next, we draw the edges $e_1,\dots,e_4$ inside $\mathcal{R}(\mu)$ such that $e_1$ crosses $e_4$ while $e_2,e_3$ are drawn crossing-free. Moreover, $e_1$ is drawn at slope $1$ and $e_4$ at slope $-1$. If the skeleton of $\nu_1$ is not empty after its modification, we insert a vertical virtual edge $e=(v_1,v_2)$ for the remainder of $\nu_1$ and define the reserved region $\mathcal{R}(\nu_1)$ which is free as it is a subset of $\mathcal{R}(\mu)$. Similarly, we add a vertical virtual edge $e'=(v_3,v_4)$ if $skel(\nu_2)$ is non-empty and define its reserved region $\mathcal{R}(\nu_2)$ which again is free as it also is a subset of $\mathcal{R}(\mu)$.  See~\cref{fig:subcubic-p-2} for the construction. Otherwise, the crossing-free edges $e_2,e_3$ are drawn as shown in~\cref{fig:subcubic-p-3} inside $\mathcal{R}(\mu)$. The inserted virtual edges and the respective reserved regions clearly fulfill \ref{inv:sc-ve-vertical} and~\ref{inv:sc-res-region}, while all other drawn edges correspond to real edges in $G$ and are not considered in the subsequent processing of $\mathcal{T}$. The only crossing is the one of $e_1$ and $e_4$, which maintains \ref{inv:sc-45deg}. Due to this crossing, all vertices are incident to the outer cell, guaranteeing \ref{inv:sc-outer}.

\begin{figure}[t]
    \begin{subfigure}[b]{.21\textwidth}
    \centering    
    \includegraphics[width=\linewidth,page=4]{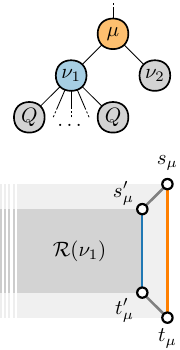}
    \subcaption{}
    \label{fig:subcubic-p-1}
    \end{subfigure}
    \hfil
    \begin{subfigure}[b]{.21\textwidth}
    \centering
    \includegraphics[width=\linewidth,page=5]{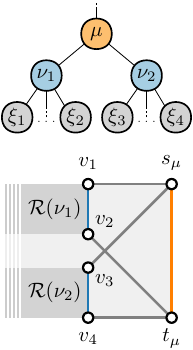}
    \subcaption{}
    \label{fig:subcubic-p-2}
    \end{subfigure}
    \hfil
    \begin{subfigure}[b]{.21\textwidth}
    \centering
    \includegraphics[width=\linewidth,page=6]{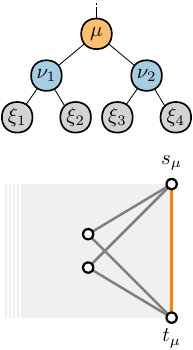}
    \subcaption{}
    \label{fig:subcubic-p-3}
    \end{subfigure}
    \hfil
    \begin{subfigure}[b]{.24\textwidth}
    \centering
    \includegraphics[width=\linewidth,page=7]{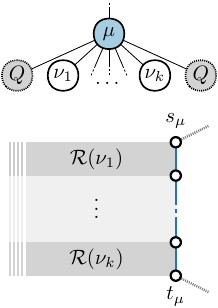}
    \subcaption{}
    \label{fig:subcubic-s}
    \end{subfigure}
    \caption{(a - c) Treatment of $P$-nodes in the algorithm in the proof of \cref{thm:sp-degree-3}. (d) Treatment of $S$-nodes in the algorithm in the proof of \cref{thm:sp-degree-3}.}
    \label{fig:subcubic-p}
\end{figure}

\paragraph{$\mu$ is a $S$-node.}  
    $skel(\mu)$ consists of a path of virtual edges from $s_\mu$ to $t_\mu$. $s_\mu$ and $t_\mu$ were already placed when processing the parent node in $\mathcal{T}$ such that $(s_\mu,t_\mu)$ is a vertical segment by \ref{inv:sc-ve-vertical}. Let $e_i = (u_i,v_i)$ be the virtual edge corresponding to child node $\nu_i$. We draw $e_1,\dots,e_k$ as equally sized, consecutive vertical lines between $s_\mu$ and $t_\mu$, maintaining \ref{inv:sc-ve-vertical} and~\ref{inv:sc-outer}. It follows that $\mathcal{R}(\mu)$ which is free by \ref{inv:sc-res-region} is vertically divided into $k$ equally sized sub-regions; see~\cref{fig:subcubic-s}. As the reserved regions $\mathcal{R}(\nu_1),\dots,\mathcal{R}(\nu_k)$ only share common borders and are located inside $\mathcal{R}(\mu)$, \ref{inv:sc-res-region} is guaranteed. As there are no crossings, \ref{inv:sc-45deg} is maintained.

\paragraph*{Correctness.} For a parent virtual edge corresponding to a not yet drawn node $\mu$, the endpoints  are the poles $s_\mu, t_\mu$. Thus,  \ref{inv:sc-ve-vertical} ensures that the reserved region $\mathcal{R}(\mu)$ is well defined. Further, \ref{inv:sc-res-region} guarantees that there is no overlap between different parts of the drawing, as the skeleton of each node $\mu$ is drawn within $\mathcal{R}(\mu)$ and each child of $\mu$ gets a free area that is part of $\mathcal{R}(\mu)$ as described above. Since reserved regions of sibling nodes do not overlap, all occurring crossings in $\Gamma$ are drawn explicitly in our algorithm and \ref{inv:sc-45deg} ensures that they are right-angled and have the same slope.  Finally, all vertices in $G$ are incident to the outer face by \ref{inv:sc-outer}. As \ref{inv:sc-ve-vertical}~to~\ref{inv:sc-outer} are maintained, the resulting drawing is outer-apRAC.

\paragraph{Running time.} We construct  $\mathcal{T}$ in linear time due to~\cite{DBLP:conf/gd/GutwengerM00}. As there are $\mathcal{O}(1)$ operations per node, the top-down visit is in $\mathcal{O}(n)$, resulting in $\mathcal{O}(n)$ time.
\qed\end{proof}

Using similar but more sophisticated techniques we prove in \cref{app:degree4shit}:

\begin{restatable}
{theorem}{spdegreefour}\label{thm:sp-degree-4}
Let $G$ be a biconnected SP-graph with maximum degree $4$. An outer-RAC drawing of $G$ can be computed in $\mathcal{O}(n)$ time.
\end{restatable}

\section{Open Problems}

We conjecture that all subcubic planar graphs  and  all subquartic SP-graphs are outer-apRAC. For the first conjecture, one needs to draw triconnected subcubic planar graphs whereas for the second one we need a technique for S-nodes that introduces crossings in certain cases.
Also, we believe that a high vertex degree at any vertex is also an obstruction for outer-RAC; such a property may be useful for a full characterization. 
Moreover, an efficient recognition algorithm for general SP- or planar graphs is of interest as well as an area-efficient drawing algorithm for subquartic SP-graphs. To this end note that our drawing algorithms produce exponential-area drawings.
Finally, \cref{thm:planar} motivates to study outer-RAC drawings where edges have bends or are drawn with circular arcs~\cite{DBLP:conf/swat/ChaplickFK020}. To this end, also note that every graph is outer-apRAC with three bends per edge~\cite{DBLP:journals/tcs/DidimoEL11}. 

\bibliographystyle{splncs04}
\bibliography{rac}

\appendix
\newpage
\section{Proof of \cref{thm:density-upperbound}}
\label{app:densityshit}
\densityshit*
\begin{proof}
    Consider any outer-RAC drawing $\Gamma$ of $G$. We assume w.l.o.g. that $\Gamma$ is connected, otherwise, we can connect the single components in a treelike structure by adding edges. We extend $G$ by all the outlines of blocks and then triangulate all faces bounded by crossing-free edges to obtain a drawing $\Gamma'$ of a supergraph $G'$ of $G$ where outlines are interpreted as cycles of edges. This may introduce parallel edges and self-loops. We will show now that the cycle formed by two parallel edges $e$ and $e'$ is necessarily empty of other edges and vertices, i.e., we can remove $e$ or $e'$ from $G'$. To this end, note that at least one of $e$ and $e'$, say $e'$ must be part of an outline boundary. Assume for a contradiction that $e \in B$ for some block $B$. Since $e \in B$, we have that $e=(u,v)$ is crossed by an edge $e^*=(u^*,v^*)$ of $B$ such that $u,u^*,v,v^*$ occur in this order along the outer boundary $C$ of $\Gamma$. Since $e'$ is located inside $C$, it must cross $e^*$, a contradiction. Thus, we have $e \in O(B)$ and $e'\in O(B')$ for two distinct blocks $B, B'$. Now, assume for a contradiction that the bounded region $R$ bounded by the cycle formed by $e$ and $e'$ was not empty. Since $e$ and $e'$ are crossing-free, $R$ must contain at least one block $B^*$ with all of its endpoints, a contradiction to the fact that $\Gamma$ is outer-RAC. 
    By the same argument, we can argue that self-loops are empty and hence we can remove them. 
    This guarantees that at most two blocks $B$ and $B'$ share an outline edge. 
    
    We prove that even $G'$ has the required edge density. Note that $G'$ still has the property that edges cross only at right angles. Moreover, each block is bounded by a cycle of plane edges that might be drawn non-straight-line. Hence, each block together with its outline induces a bounded block subgraph. Moreover, we call faces bounded by only crossing-free edges \emph{trivial bounded blocks} and we call single edges on the outer cycle \emph{degenerate bounded blocks}. Finally, we call bounded blocks that are neither trivial nor degenerate \emph{proper}. Since the outer cycle still contains all vertices, we conclude that we can find a sequence of bounded blocks $B_0,B_1,\ldots,B_k$, such that starting from $B_0$ we can construct $\Gamma'$ by inserting $B_1,\ldots,B_k$ in order such that $B_i$ shares either one edge or one vertex with one of the previous bounded blocks $B_0,B_1,\ldots,B_{i-1}$ for $i > 1$. Note that we can assume w.l.o.g. that $B_0$ is degenerate.

    For block $B_i$, let $n_i$ and $m_i$ denote its number of vertices and its number of edges, respectively.  If $B_i$ is trivial, we have $n_i=m_i=3$. If $B_i$ is degenerate, we have $n_i=2$ and $m_i=1$. Finally, if $B$ is proper, we have $m_i \leq 2n-2$ by \cref{lem:boundedblock}. Now, let $G_i'$ denote the subgraph of $G'$ consisting of subgraphs $B_1,\ldots,B_i$. Let $n_i'$ and $m_i'$ denote the difference in the number of vertices and edges  of $G_i$ and $G_{i-1}$, respectively. We distinguish five cases:
    \begin{enumerate}
        \item If $B_i$ is degenerate, it shares a single vertex with $G_{i-1}$, $m_i' = 1 = n_i' < 2n_i' + 1$.
        \item If $B_i$ is trivial and shares a single vertex with $G_{i-1}$, $m_i' = 3 = \frac{3}{2} n_i' < 2n_i' + 1$.
        \item If $B_i$ is trivial and shares a single edge with $G_{i-1}$, $m_i' = 2 = 2 n_i' < 2n_i' + 1$.
        \item If $B_i$ is proper and shares a single vertex with $G_{i-1}$, $m_i' \leq 2 (n_i'+1)-2 = 2n_i' < 2n_i' + 1$.
        \item If $B_i$ is proper and shares a single edge with $G_{i-1}$, $m_i' \leq 2 (n_i'+2)-3 = 2n_i' + 1$.
    \end{enumerate}
    Let $k_p$ denote the number of times we insert a proper block and let $k_t$ denote the number of times we insert a trivial or degenerate block. Let $n_i^p$ and $m_i^p$ denote the number of vertices and edges of the $i$-th proper block inserted, respectively, and let $n_i^t$ and $m_i^t$ denote the corresponding values for the $i$-th trivial or degenerate block. If we insert the $i$-th proper block, by the above analysis, we have $m_i^p \leq 2n_i^p + 1$, if we insert the $i$-th trivial or degenerate block, we have $m_i^t \leq 2n_i^t$. We can now sum up the number of vertices $n'$ and edges $m'$ of $G'$:
    \begin{align}
    n'&=2 + \sum_{i=1}^{k_p}n_i^p + \sum_{i=1}^{k_t}n_i^t\\
    m'&\leq1 + \sum_{i=1}^{k_p} \left(2n_i^p+1\right) + \sum_{i=1}^{k_t} \left(2n_i^t\right) =(k_p+1) + 2\sum_{i=1}^{k^p}n_i^p+2\sum_{i=1}^{k^t}n_i^t\nonumber\\
    &\label{eqn:totaldensity}=(k_p-3) + 2\left(2 + \sum_{i=1}^{k_p}n_i^p + \sum_{i=1}^{k_t}n_i^t\right)=(k_p-3)+2n'
    \end{align}
    Since each proper block has at least $4$ vertices and the initial degenerate block has $2$ vertices,
    \begin{equation}
    \label{eqn:k_p}   k_p \leq \frac{n'-2}{2}=0.5n'-1.
    \end{equation}
    Inserting \eqref{eqn:k_p} into \eqref{eqn:totaldensity} now yields $m' \leq 2.5n'-4$.

    If we want to tightly achieve $m'=2.5n'-4$, we must obtain equality for Inequalities~\eqref{eqn:totaldensity} and \eqref{eqn:k_p}. For \eqref{eqn:totaldensity}, this is the case if each $B_i$ is proper and shares a single edge with $G_{i-1}$. For  \eqref{eqn:k_p}, we achieve equality exactly if each proper block has $4$ vertices, i.e., it is a $K_4$.
\qed\end{proof}
\section{Required Crossings in Skeletons of S-nodes in Outer-apRAC Drawings}
    \label{app:necessarycrossingshit}
    \begin{figure}[t]
    \centering
    \begin{minipage}{0.7\textwidth}
    \centering
    \begin{subfigure}{0.4\textwidth}
        \centering
        \includegraphics[page=3]{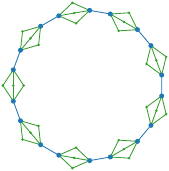}
        \subcaption{}
        \label{fig:l9-structure}
    \end{subfigure}
    \hfil
    \begin{subfigure}{0.4\textwidth}
        \centering
        \includegraphics[page=4]{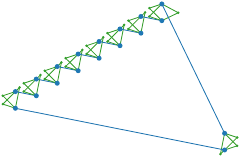}
        \subcaption{}
        \label{fig:l9-outerRAC}
    \end{subfigure}
  
    \begin{subfigure}{0.4\textwidth}
        \centering
        \includegraphics[page=5]{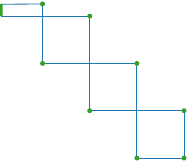}
        \subcaption{}
        \label{fig:l9-cycle}
    \end{subfigure}
    \hfil
    \begin{subfigure}{0.4\textwidth}
        \centering
        \includegraphics[page=6]{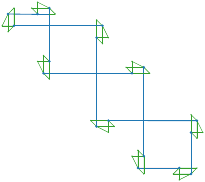}
        \subcaption{}
        \label{fig:l9-outerapRAC}
    \end{subfigure}
    \end{minipage}
    \hfil
    \begin{minipage}{0.25\textwidth}
    \begin{subfigure}{1\textwidth}
        \centering
        \includegraphics[page=7]{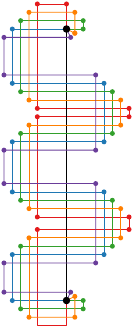}
        \subcaption{}
        \label{fig:highdegree}
    \end{subfigure}      
    \end{minipage}
    \caption{(a)~Graph $L_9 \in \mathcal{L}$. (b)~Outer-RAC drawing of $L_9$. (c)~Drawing of a $9$-cycle as described in the proof of \cref{lem:not-nice-ap}. (d)~Outer-apRAC drawing of $L_9$. (e)~An outer-apRAC SP-graph with maximum degree $6$.}
    \label{fig:l9}
\end{figure}

    We define the graph family $\mathcal{L}$ called \emph{linked $K_{2,3}$'s}. For every $k \geq 2$, there is a graph $L_k \in \mathcal{L}$. More precisely, $L_k$ consists of $k$ graphs $G_1,\ldots,G_k$ isomorphic to $K_{2,3}$ --- i.e., $G_i$ consists of the three paths $(s_i,x_i,t_i)$, $(s_i,y_i,t_i)$ and $(s_i,z_i,t_i)$ --- as well as $k$ \emph{links}, that is, edges $(t_i,s_{i+1})$ for $1 \leq i \leq k$  (indices taken modulo $k$); see \cref{fig:l9-structure} for an illustration of $L_9$. In particular, in the SPQR-tree decomposition $L_k$ contains a node $\sigma_k$ whose skeleton is the cycle $(s_1,t_1,s_2,t_2,\ldots,s_k,t_k)$. First, observe the following; see \cref{fig:l9-outerRAC}:
    
    \begin{observation}
        For $k \geq 2$, $L_k$ admits an outer-RAC drawing where the skeleton of $\sigma_k$ is crossing-free.
    \end{observation}

    On the other hand, if we require an outer-apRAC drawing, $\sigma_k$ may be non-planar.
    
    \begin{restatable}{theorem}{notniceap}\label{lem:not-nice-ap}
        If $k \geq 9$, in any outer-apRAC drawing of $L_k$, the skeleton of $\sigma_k$ has crossings.
        Further, for $k \geq 2$, $L_k$ admits an outer-apRAC drawing.
    \end{restatable}
    \begin{proof}
        For the first part of the theorem, assume for a contradiction, that $k \geq 9$ and that there exists an outer-apRAC drawing of $L_k$ where $\sigma_k$ has no crossings. In the following discussion, all indices are taken modulo $k$. By \cref{lem:three-parallel-short-paths}, the topology of each subgraph $G_i$ is uniquely defined. That is, along the outer cycle, $z_i$ is separated from $x_i$ and $y_i$ by vertices $s$ and $t$. We first claim the following:
        
        \begin{myclaim}\label{clm:no-other-edge-crosses}
            $G_i$ cannot be crossed by any edge $e$ except for $(t_i,s_{i+1})$ and $(s_i,t_{i-1})$.
        \end{myclaim}
        \begin{proof}Assume for a contradiction that such an edge $e$ existed. Clearly, both its endpoints must be located on the outer face. Thus, $e$ must cross $G_i$ at least twice. By \cref{lem:three-parallel-short-paths}, it can only cross edges incident to $z_i$. However, these share an endpoint and thus are not parallel; a contradiction.
        \qed\end{proof}

        Next, we show the following:
        \begin{myclaim}\label{clm:one-link-crosses}
           Either $(t_i,s_{i+1})$ crosses $(s_i,z_i)$ or that $(s_i,t_{i-1})$ crosses $(t_i,z_i)$.  
        \end{myclaim}

        \begin{proof}
        Assume for a contradiction that this was not the case. 
        First, assume that $(t_i,s_{i+1})$ crosses $(s_i,z_i)$ and that  $(s_i,t_{i-1})$ crosses $(t_i,z_i)$. Then, in order to preserve $z_i$'s visibility to the outer cycle, we also must have that $(t_i,s_{i+1})$  crosses  $(s_i,t_{i-1})$; a contradiction as now $(t_i,s_{i+1})$ must cross two edges incident to $s_{i+1}$ at a right angle.          
        Thus, we must have that neither $(t_i,s_{i+1})$ crosses $(s_i,z_i)$ nor that $(s_i,t_{i-1})$ crosses $(t_i,z_i)$. In this scenario, by \cref{clm:no-other-edge-crosses}, the path between $t_i$ and $s_i$  via $s_{i+1}$ and $t_{i-1}$ encloses at least one vertex of $G_i$; a contradiction.
        \qed\end{proof}

        Next, we prove that consecutive links have distinct slopes.

        \begin{myclaim}\label{clm:angle-at-k23}
            The angle formed by $(t_i,s_{i+1})$  and $(t_{i-1},s_i)$ inside the outer cycle is less than $\pi$.
        \end{myclaim}
        \begin{proof}
            W.l.o.g. assume that $(t_i,s_{i+1})$ crosses the path $(s_i,z_i,t_i)$ (the case where $(t_{i-1},s_i)$ crosses $(s_i,z_i,t_i)$ is symmetric by \cref{clm:one-link-crosses}).
            Let $c_1$ denote the crossing between $(s_i,x_i,t_i)$ and $(s_i,y_i,t_i)$ and $c_2$ denote the crossing between $(s_i,z_i,t_i)$ and and $(t_i,s_{i+1})$. Consider the quadrangle $Q=\Box s_ic_1t_ic_2$. Since $c_1$ and $c_2$ form an angle of $\pi$ in $Q$, the angles at $t_i$ and $s_i$ must be convex. This is only possible, if $(t_i,s_{i+1})$ and $(s_i,c_1)$ are parallel. Then,   $(t_{i-1},s_i)$ must be drawn in the wedge between $s_ic_1$ and $s_ic_2$ outside of $Q$ by \cref{clm:one-link-crosses}; i.e., it is not parallel to $(t_i,s_{i+1})$. Since the drawing of the skeleton of $\sigma_k$ is crossing-free, the convex angle is inside the outer cycle.  
        \qed\end{proof}

        We are now ready to prove the first statement of the theorem. Since every subgraph $G_i$ is incident to a link involved in crossings by \cref{clm:one-link-crosses}, at least $\lceil \frac{k}{2} \rceil \geq 5$ links are crossing. By \cref{clm:angle-at-k23}, each  consecutive pair of crossing links along the skeleton of $\sigma_k$ must have different rotation in a cyclic walk along the skeleton of $\sigma_k$. Since there are only four such rotations for links in apRAC, we arrive at a contradiction. 

        For the second part of the statement consider the following drawing algorithm. We first draw $\sigma_k$ recursively as follows in such a way, that  each vertex $v$ is represented by a point with the cycle forming an interior angle of $\pi/2$ at $v$ or by a vertical segment forming an interior angle of $\pi$ with the cycle  at $v$. A $2$-cycle consists of two  vertices represented by vertical segments and two horizontal edges, i.e., the interior angle at each vertex is $\pi$. Recursively, if we draw a $k$-cycle, we first draw a $k-1$-cycle. Then, if there is a vertex with an interior angle of $\pi$ which is drawn as a vertical segment, we replace it by two vertices drawn as points and a vertical edge connecting them. Otherwise, we pick any vertex  $v$ represented by a point. We remove $v$ and extend the two edges meeting at $v$ so to form a proper crossing. Then, we insert a vertex $v_1$ represented by a point at the end of the vertical edge and a vertex $v_2$ represented by a vertical segment at the end of the horizontal edge and connect $v_1$ and $v_2$ with a horizontal edge; see \cref{fig:l9-cycle} for a $9$-cycle drawn this way. 
        In the resulting representation, each edge will correspond to one of the links and each vertex to one of the $K_{2,3}$. We now replace each vertex by a $K_{2,3}$ as exemplified in \cref{fig:l9-outerapRAC}. The theorem follows.
    \qed\end{proof}

  Finally, we remark that there also exist SP-graphs of higher degree that admit outer-RAC drawings; see e.g. \cref{fig:highdegree}. 

\section{Outer-RAC Drawings of Subquartic SP-Graphs}
\label{app:degree4shit}

\spdegreefour*
\begin{proof}
    Let $G$ be a subquartic SP-graph and $\mathcal{T}$ its SPQR-tree. It follows that $\mathcal{T}$ contains no $R$-nodes and $P$-nodes have either two or three children. We construct the outer-RAC drawing $\Gamma$ again in a top-down fashion, traversing $\mathcal{T}$ pre-order from a root node that we select using \cref{prop:subcubic-root}. When processing a node $\mu$ in $\mathcal{T}$, 
    we draw the virtual edges in $skel(\mu)$, thus the parent virtual edge  is already drawn when processing a node $\mu$. As a result, if we process $\mu$, its poles $s_\mu$ and $t_\mu$ have already been assigned positions fixing the drawing of the parent virtual edge $(s_\mu,t_\mu)$ and we mostly draw the remainder of the skeleton of $\mu$ in a \emph{reserved region $\mathcal{R}(\mu)$} around $(s_\mu,t_\mu)$. The definition of region  $\mathcal{R}(\mu)$ is more involved compared to \cref{thm:sp-degree-3} and depends on the type of $\mu$ and the number of its children. Moreover, similar as in~\cref{thm:sp-degree-3}, $S$-nodes will be partially drawn when processing their parent node. In addition, in some cases we additionally make use of an \emph{empty region} $\mathcal{E}(\mu)$ that is not part of the outer face but can be traversed by an edge of the skeleton of $\mu$.

    \begin{figure}[t]
\centering
\begin{subfigure}{.4\textwidth}
    \centering    
    \includegraphics[scale=1,page=8]{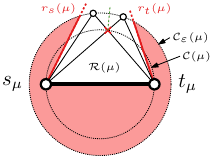}
    \subcaption{}
    \label{fig:subquartic-circles-1}
\end{subfigure}
\hfil
\begin{subfigure}{.4\textwidth}
    \centering    
    \includegraphics[scale=1,page=9]{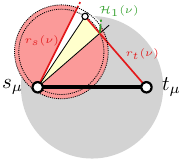}  
    \subcaption{}
    \label{fig:subquartic-circles-2}
\end{subfigure}
    \caption{(a)~Illustration for $\mathcal{R}(\mu)$ in \cref{thm:sp-degree-4}. (b)~For child node $\nu$, we have $\mathcal{R}(\nu) \subset \mathcal{R}(\mu)$.}
    \label{fig:subquartic-circles}
\end{figure}

    We define $\mathcal{R}(\mu)$ and $\mathcal{E}(\mu)$.
    Let $\mathcal{C}(\mu)$ be the circle whose diameter is the parent virtual edge $(s_\mu,t_\mu)$ of $\mu$. Moreover, let $\mathcal{C}_\varepsilon(\mu)$ be a circle concentric with $\mathcal{C}(\mu)$ but with the radius being $\varepsilon > 0$ larger than $\mathcal{C}(\mu)$. To this end, note that $\varepsilon$ is not a constant throughout the algorithm, instead we choose a suitable value when determining $\mathcal{R}(\mu)$ and $\mathcal{E}(\mu)$ in our algorithm. Observe that according to Thales's theorem, we can position a crossing between one edge incident to $s_\mu$ and one incident to $t_\mu$ anywhere on $\mathcal{C}(\mu)$ to ensure it is right-angled. The endpoints of the crossing edges can then be positioned inside $\mathcal{C}_\varepsilon(\mu)\setminus \mathcal{C}(\mu)$. Observe that as long as there is some free space on $\mathcal{C}$, we can always find a suitable $\mathcal{C}_\varepsilon(\mu)$ to insert such two crossing edges; see~\cref{fig:subquartic-circles}. The reserved region $\mathcal{R}(\mu)$ is a part of  $\mathcal{C}_\varepsilon(\mu)$ delimited by the edge $(s_\mu,t_\mu)$ (included) and two rays $r_s(\mu)$ and $r_t(\mu)$ emanating from $s_\mu$ and $t_\mu$ (both excluded), respectively; see~\cref{fig:subquartic-circles}. Moreover, we further restrict $\mathcal{R}(\mu)$ by intersection with two open halfplanes $\mathcal{H}_1(\mu)$ and $\mathcal{H}_2(\mu)$ so to avoid overlaps with other reserved regions; see see~\cref{fig:subquartic-circles} for an example where one halfplane restricts $\mathcal{R}(\mu)$ and the other does not. We do not have any strict requirements on the slopes of $r_s(\mu)$ and $r_t(\mu)$ or on $\mathcal{H}_1(\mu)$ and $\mathcal{H}_2(\mu)$ aside from the fact that $\mathcal{R}(\mu)$ must contain a segment of the boundary of $\mathcal{C}_\varepsilon(\mu)$.  
    
    \begin{figure}[t]
\centering
\begin{subfigure}{.4\textwidth}
    \centering    
    \includegraphics[scale=1,page=10]{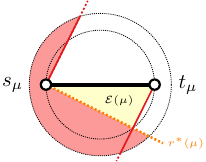}
    \subcaption{}
    \label{fig:subquartic-e-1}
\end{subfigure}
\hfil
\begin{subfigure}{.4\textwidth}
    \centering    
    \includegraphics[scale=1,page=11]{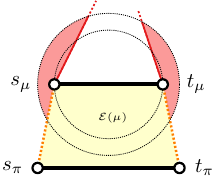}  
    \subcaption{}
    \label{fig:subquartic-e-2}
\end{subfigure}
    \caption{Illustration for $\mathcal{E}(\mu)$ in the case where $\mathcal{E}(\mu)$ is (a)~triangle-shaped and (b)~trapezoid-shaped. (a) also shows \ref{inv:4:region:ep} and \ref{inv:4:region:es1} and (b) shows \ref{inv:4:region:es2}.}
    \label{fig:subquartic-e}
\end{figure}

    In addition, for some of the nodes we will define $\mathcal{E}(\mu)$ which may have one of two shapes. Note that if  $\mathcal{E}(\mu) \neq \emptyset$, we also have that $\mathcal{E}(\mu) \cap \mathcal{R}(\mu) = \emptyset$ in both cases. If $\mathcal{E}(\mu)$ is \emph{triangle-shaped}, it is a triangle defined by edge $(s_\mu,t_\mu)$ (excluded), ray $r_t(\mu)$ (excluded) and a ray $r^*(\mu)$  perpendicular to $r_t(\mu)$ emanating from $s_\mu$ (sometimes included, sometimes excluded); see~\cref{fig:subquartic-e-1}. Note that this requires that the angle between $(s_\mu,t_\mu)$ and $r_t(\mu)$ inside $\mathcal{E}(\mu)$ is acute. Moreover, $\mathcal{E}(\mu)$ can be \emph{trapezoid-shaped}. In this scenario, $\mathcal{E}(\mu)$ is a trapezoid defined by $(s_\mu,t_\mu)$ and the edge $(s_\pi,t_\pi)$ between the nodes of the parent $\pi$ of $\mu$ which is parallel to $(s_\mu,t_\mu)$ (all boundaries are excluded); see~\cref{fig:subquartic-e-2}. Observe that in this scenario, the angles inside $\mathcal{E}(\mu)$ at $s_\mu$ and $t_\mu$ are not necessarily acute.
    We are now ready to define the invariants of our algorithm:

\begin{enumerate}[label={I.\arabic*},left=\labelsep]
\item\label{inv:4:region:r} Let $\mu$ be a not yet processed non-$Q$-node whose parent node in $\mathcal{T}$ has been processed already. The reserved region $\mathcal{R}(\mu)$ is part of the outer cell.
\item\label{inv:4:region:rnooverlap} Let $\mu_1$ and $\mu_2$ be two not yet processed non-$Q$-nodes. Then, $\mathcal{R}(\mu_1) \cap \mathcal{R}(\mu_2) = \emptyset$. 
\item\label{inv:4:region:ep} Let $\mu$ be a not yet processed $P$-node with three children. Then,  $\mathcal{E}(\mu) \neq \emptyset$ is triangle-shaped, does not contain any already drawn edges and the segment of $r^*(\mu)$ between $s_\mu$ and $r_t(\mu)$ is part of $\mathcal{E}(\mu)$.  Moreover, the edge incident to $t_\mu$ in the parent component is aligned with $r_t(\mu)$. 

\item\label{inv:4:region:es1} Let $\mu$ be a not yet processed $S$-node for which one $Q$-node child has already been drawn. Then,  $\mathcal{E}(\mu) \neq \emptyset$ is triangle-shaped,  does not contain any already drawn edges and $r^*(\mu)$  is not part of $\mathcal{E}(\mu)$. Moreover, $s_\mu$ is incident to an edge that is drawn as part of $r^*(\mu)$ and belongs to a sibling node of $\nu$. Finally, the drawn $Q$-node child is incident to $t_\mu$ and its contained edge is aligned with $r_t(\mu)$. 
\item\label{inv:4:region:es2} Let $\mu$ be a not yet processed $S$-node for which two $Q$-node children have already been drawn. Then,  $\mathcal{E}(\mu) \neq \emptyset$ is trapezoid-shaped and does not contain any already drawn edges. Moreover, the drawn $Q$-node children  are $(s_\mu,s_\pi)$ and $(t_\mu,t_\pi)$ where $\pi$ is the parent node of $\mu$. The  edges contained in the drawn $Q$-node children are aligned with the boundary of $\mathcal{E}(\mu)$. 
\item\label{inv:4:rightangle} All crossings are right-angled.
\item\label{inv:4:outer} Every already drawn vertex is incident to the outer cell. 
\end{enumerate}

\paragraph{Correctness.} For a virtual edge corresponding to a not yet drawn node $\nu$ of $\mu$,  \ref{inv:4:region:r} ensures that the reserved region $\mathcal{R}(\nu)$ has access to the outer face. Further, \ref{inv:4:region:rnooverlap} guarantees that there is no overlap between different parts of the drawing, as the skeleton of each node $\mu$ is drawn within $\mathcal{R}(\mu)$ and each child $\nu$ of $\mu$ will be assigned new areas $\mathcal{R}(\nu)$ and $\mathcal{E}(\nu)$. In the drawing we will make use of \ref{inv:4:region:ep} to~\ref{inv:4:region:es2} as discussed below. All occurring crossings in $\Gamma$ are drawn explicitly and \ref{inv:4:rightangle} ensures that they are right-angled.  Finally, all vertices in $G$ are incident to the outer face by \ref{inv:4:outer}.  Thus, if \ref{inv:4:region:r}~to~\ref{inv:4:outer} are maintained, the resulting drawing is outer-RAC.
    

We now first describe how to compute an initial partial drawing adhering to \ref{inv:4:region:r}--\ref{inv:4:outer} and then describe how we handle the different node types in the top-down traversal. 

\paragraph{Initialization.}  Observe that if $\mathcal{T}$ contains no $P$-node, $G$ is a cycle and therefore outerplanar and thus also outer-RAC. Hence in the following, we can assume that $G$ contains at least one $P$-node. Then we root $\mathcal{T}$ according to~\cref{prop:subcubic-root}. Now, the root $\mu_r$ has as a child a $S$-node $\mu_s$ which in turn has exactly one $P$-node child $\mu_p$. In $G$, according to the skeleton of $\mu_s$, we have that $\mu_r$ and the $Q$-node children of $\mu_s$ form a path $P$ connecting the poles of $\mu_p$. As $P$ contains at least two edges ($\mu_r$ and at least one $Q$-node child of $\mu_s$) we draw the path $P$ such that the poles of $\mu_p$ are vertically aligned and that the angle at $t_{\mu_p}$ is acute as depicted in~\cref{fig:subquartic-init}. Clearly, we have that \ref{inv:4:region:r} and~\ref{inv:4:region:rnooverlap} hold for $\mu_p$. As there are no crossings and the drawing is outer plane, \ref{inv:4:rightangle},  and~\ref{inv:4:outer} are guaranteed. Finally, since the drawing of the cycle is empty, we maintain \ref{inv:4:region:ep} for $\mu_p$ whereas \ref{inv:4:region:es1} and~\ref{inv:4:region:es2} are trivially fulfilled. This concludes the base case of the algorithm.

\begin{figure}[t]
    \centering    
    \includegraphics[scale=1,page=12]{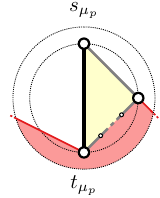}
    \caption{Initialization of the algorithm of \cref{thm:sp-degree-4}.}
    \label{fig:subquartic-init}
\end{figure}

\paragraph{$\mu$ is a $Q$-node.} The single edge in $skel(\mu)$ which corresponds to an edge in $G$ is simply drawn as a segment between the already positioned endpoints $s_\mu$ and $t_\mu$ coinciding with the already drawn parent virtual edge of $\mu$. 

\begin{figure}[t]
\centering
    \begin{subfigure}[b]{.32\textwidth}
    \centering    
    \includegraphics[page=13]{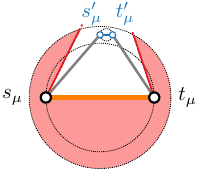}
    \subcaption{}
    \label{fig:sp-4-p-1}
    \end{subfigure}
    \hfil
    \begin{subfigure}[b]{.32\textwidth}
    \centering
    \includegraphics[page=14]{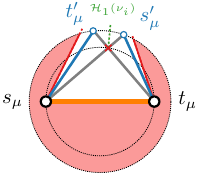}
    \subcaption{}
    \label{fig:sp-4-p-2}
    \end{subfigure}
    \hfil
    \begin{subfigure}[b]{.32\textwidth}
    \centering
    \includegraphics[page=15]{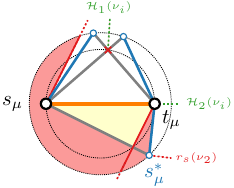}
    \subcaption{}
    \label{fig:sp-4-p-3}
    \end{subfigure}  
    \caption{Treatment of $P$-nodes in the algorithm in the proof of \cref{thm:sp-degree-4}.}
    \label{fig:sp-4-p}
\end{figure}

\paragraph*{$\mu$ is a $P$-node with one $S$-node child.} \label{p-node-1}
 Due to simplicity, $\mu$ can only have one $Q$-node child. Thus, let $\nu_1$ be the $S$-node child and $\nu_2$ be the $Q$-node child of $\mu$. Further, let $e_s = (s_\mu, s'_\mu)$ and $e_t = (t_\mu, t'_\mu)$ be the edges in $skel(\nu_1)$ incident to the vertices coinciding with the poles of $\mu$. We partially draw $skel(\nu_1)$, by placing $s'_\mu$ and $t'_\mu$. Then, we remove $e_s$ and $e_t$ from $skel(\nu_1)$ and define $s'_\mu$ and $t'_\mu$ as the new poles of  $\nu_1$. By \ref{inv:4:region:r}, there is a  region $R$ inside $\mathcal{C}_\varepsilon(\mu) \setminus \mathcal{C}(\mu)$ that is part of the outer cell. Within $R$, we place $s'_\mu$ and $t'_\mu$ close to the intersection of  $r_s(\mu)$ with $\mathcal{C}_\varepsilon$ such that $(s'_\mu,t'_\mu)$ is parallel to $(s_\mu,t_\mu)$; see~\cref{fig:sp-4-p-1}. Moreover, in this process, we ensure that $\mathcal{R}(\nu_1)$ is entirely contained within $\mathcal{R}(\mu)$. This guarantees \ref{inv:4:region:r} and~\ref{inv:4:region:rnooverlap}. Moreover, since $(s'_\mu,t'_\mu)$ and $(s_\mu,t_\mu)$ are parallel, we yield \ref{inv:4:region:es2}. Finally, no new crossings are created and $s'_\mu$ and $t'_\mu$ are placed inside the outer cell and we obtain \ref{inv:4:rightangle} and~\ref{inv:4:outer}. For $\nu_2$ we do not have to do anything as its parent virtual edge is $(s_\mu,t_\mu)$. As we do not insert any other virtual edge corresponding to $S$- or $P$-nodes, we also satisfy \ref{inv:4:region:ep} and~\ref{inv:4:region:es1}.

\paragraph*{$\mu$ is a $P$-node with two $S$-node children.} 
Let $\nu_1$ and $\nu_2$ be the two $S$-node children of $\mu$. Note that $\mu$ can have another $Q$-node child, which will be drawn between the poles of $\mu$. Due to $G$ being subquartic, there are at least two edges $e_1 \in skel(\nu_1)$ and $e_2 \in skel(\nu_2)$ corresponding to $Q$-nodes where w.l.o.g. $e_1=(s_\mu,s'_\mu)$ is incident to $s_\mu$ and $e_2=(t_\mu,t'_\mu)$ to $t_\mu$ (the other case is symmetric). We will draw $s'_\mu$ and $t'_\mu$ in this step and delete $e_1$ from $skel(\nu_1)$ and $e_2$ from $skel(\nu_2)$ making $s'_\mu$ and $t'_\mu$ the new poles of $\nu_1$ and $\nu_2$, respectively. By \ref{inv:4:region:r}, there is a  region $R$ inside $\mathcal{C}_\varepsilon(\mu) \setminus \mathcal{C}(\mu)$ that is part of the outer cell. We draw $e_1$ and $e_2$ so that they are crossing, and place the crossing point at the center of the free interval of $\mathcal{C}$. By elongating the line from $s_\mu$ and $t_\mu$ to this point, we place the endpoints of $e_1$ and $e_2$ on $\mathcal{C}_\varepsilon$. In this process, we ensure that $\mathcal{R}(\nu_1)$ and $\mathcal{R}(\nu_2)$ are entirely contained within $\mathcal{R}(\mu)$. Moreover, we separate $\mathcal{R}(\nu_1)$ and $\mathcal{R}(\nu_2)$ along the bisector $b$ of the right angle occuring at the crossing of $e_1$ and $e_2$ to make them non-overlapping~\cref{fig:sp-4-p-2}. Note that this separation corresponds to the intersection with a half-plane $\mathcal{H}_1(\nu_i)$ delimited by $b$ for $(\nu_i)$. This guarantees \ref{inv:4:region:r} and~\ref{inv:4:region:rnooverlap}. Moreover, $e_1$ and $e_2$ cross at an right angle, hence we can make the angle in the enclosed cell at $s'_\mu$ and $t'_\mu$ acute. This construction guarantees  \ref{inv:4:region:es1}. Finally, no new crossings are created and $s'_\mu$ and $t'_\mu$ are placed inside the outer cell and we obtain \ref{inv:4:rightangle} and~\ref{inv:4:outer}. As we do not insert any other virtual edge corresponding to $S$- or $P$-nodes, we also satisfy \ref{inv:4:region:ep} and~\ref{inv:4:region:es2}.

\begin{figure}[t]
    \centering
    \begin{subfigure}{0.4\textwidth}
        \centering
        \includegraphics[page=16]{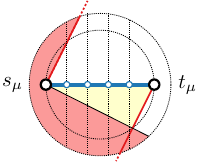}
        \subcaption{}
        \label{fig:sp-4-s-1}
    \end{subfigure}
    \hfil
    \begin{subfigure}{0.4\textwidth}
        \centering
        \includegraphics[page=17]{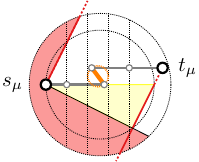}
        \subcaption{}
        \label{fig:sp-4-s-2}
    \end{subfigure}
\caption{Treatment of $S$-nodes in the algorithm in the proof of~\cref{thm:sp-degree-4}.}
    \label{fig:sp-4-s-standard}
\end{figure}

\paragraph*{$\mu$ is a $P$-node with three $S$-node children.} 
Let $\nu_1,\nu_2,\nu_3$ be the three $S$-node children of $\mu$. Since $G$ is subquartic, all virtual edges in $skel(\nu_1)$, $skel(\nu_2)$ and $skel(\nu_3)$ which are incident to the poles of $\mu$ correspond to $Q$-nodes. Let $e_1 = (s_\mu,s_\mu') \in skel(\nu_1)$, $e_2 = (s_\mu,s_\mu^*) \in skel(\nu_2)$ and $e_3 = (t_\mu,t_\mu') \in skel(\nu_3)$. We will place vertices $s_\mu'$, $s_\mu^*$ and $t_\mu'$ and remove $e_1$, $e_2$ and $e_3$ from $skel(\nu_1)$, $skel(\nu_2)$ and $skel(\nu_3)$ respectively. Then, $s_\mu'$, $s_\mu^*$ and $t_\mu'$ will be assigned the new poles of $skel(\nu_1)$, $skel(\nu_2)$ and $skel(\nu_3)$, respectively.

We draw $e_1$ and $e_3$ crossing, following the construction described in the case for $P$-nodes with two $S$-node children. Hence, it remains to discuss the drawing of  $e_2$. To this end, we use \ref{inv:4:region:ep} and draw $e_2$ following $r^*(\mu )$ crossing through $\mathcal{E}(\mu)$ and  positioning $s_\mu^*$ in the region $\mathcal{C}_\varepsilon(\mu) \setminus \mathcal{C}(\mu)$. We now can assign a dummy boundary for determining $\mathcal{R}(\nu_2)$. Moreover, we separate $\mathcal{R}(\nu_2)$ from $\mathcal{R}(\nu_1)$ and $\mathcal{R}(\nu_3)$ by the line $\ell$ containing edge $(s_\mu,t_\mu)$. Note that this separation corresponds to the intersection with a half-plane $\mathcal{H}_2(\nu_i)$ delimited by $\ell$ for $(\nu_i)$. This establishes \ref{inv:4:region:r} and~\ref{inv:4:region:rnooverlap}; see~\cref{fig:sp-4-p-3}. Since $r^*(\mu)$ and $r_t(\mu)$ are perpendicular, we maintain \ref{inv:4:rightangle}. As $s^*_\mu$ is placed on the outer face, we also maintain \ref{inv:4:outer}. Finally, the angle at $s_\mu^*$ in the enclosed cell is necessarily acute which yields \ref{inv:4:region:es1}. As we do not insert any other virtual edge corresponding to $S$- or $P$-nodes, we also satisfy \ref{inv:4:region:ep} and~\ref{inv:4:region:es2}.

\begin{figure}[t]
    \centering
    \begin{subfigure}{0.3\textwidth}
        \centering
        \includegraphics[page=18]{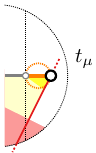}
        \subcaption{}
        \label{fig:sp-4-s-node-app-1}
    \end{subfigure}
    \hfil
    \begin{subfigure}{0.3\textwidth}
        \centering
        \includegraphics[page=19]{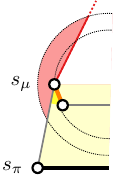}
        \subcaption{}
        \label{fig:sp-4-s-node-app-2}
    \end{subfigure}
    \hfil
    \begin{subfigure}{0.3\textwidth}
        \centering
        \includegraphics[page=20]{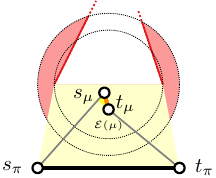}
        \subcaption{}
        \label{fig:sp-4-s-node-app-3}
    \end{subfigure}
    \caption{Treatment of $S$-nodes in the algorithm in the proof of~\cref{thm:sp-degree-4}.}
    \label{fig:sp-4-s-node-app}
\end{figure}

\paragraph*{$\mu$ is an $S$-node.} 
Let $\nu_1,\dots,\nu_k$ be the child nodes of $\mu$ in $\mathcal{T}$ and let $\nu_k$ be the $Q$-node child of $\mu$ that is already drawn. Let $e_1,\dots, e_k$ be the corresponding virtual edges which form a path. Assume that the virtual edge corresponding to $\nu_k$ is incident to $t_\mu$; the other case is symmetric.  
Initially, we will draw the virtual edges $e_1,\dots,e_{k-1}$ by uniformly subdividing $(s_\mu,t_\mu)$. Unless $\nu_i$ is a $P$-node with three $S$-node children, we can now simply assign a region $\mathcal{R}(\nu_i)$ bounded with rays through its poles perpendicular to $e_i$; see~\cref{fig:sp-4-s-1}. That way, we guarantee \ref{inv:4:region:r}, \ref{inv:4:region:rnooverlap} and trivially \ref{inv:4:region:ep} to~\ref{inv:4:outer}. Thus, it remains to consider the case where $\nu_i$ is a $P$-node with three $S$-node children.

If $\nu_i$ is a $P$-node with three $S$-node children, it follows immediately that both $\nu_{i-1}$ and $\nu_{i+1}$ must be $Q$-nodes if they exist. If both $\nu_{i-1}$ and $\nu_{i+1}$ exist, we redraw $e_{i-1},e_i,e_{i+1}$ in the staircase pattern shown in~\cref{fig:sp-4-s-2}. Note that in particular we draw $e_i$ so small, that it fulfills both \ref{inv:4:region:r} and \ref{inv:4:region:rnooverlap}. Moreover, we can make the angle between $e_i$ and $e_{i+1}$ in the interior cell acute in such a way that we fulfill \ref{inv:4:region:ep}. Following these adjustments, we have to slightly reposition $t_\mu$ which can be done by slightly elongating the $Q$-node child incident to $t_\mu$; see~\cref{fig:sp-4-s-2}.

Thus, it remains to consider the boundary cases, namely, $i \in \{1,k-1\}$. First assume that $k>2$.
If the angle at $s_\mu$ or $t_\mu$ is acute, \ref{inv:4:region:es1} or~\ref{inv:4:region:es2} already guarantees us that $\mathcal{E}(\mu)$ does not contain edges. This allows us to define region $\mathcal{E}(\nu_i)$ as a subset of $\mathcal{E}(\mu)$; see~\cref{fig:sp-4-s-node-app-1} which yields \ref{inv:4:region:ep}. Otherwise, we can redraw w.l.o.g. $e_1=(s_\mu,s_\mu')$ so that we place $s_\mu'$ further into the $\mathcal{E}(\mu)$ as shown in~\cref{fig:sp-4-s-node-app-2}. Again, we draw $e_1$ so small, that it fulfills both \ref{inv:4:region:r} and \ref{inv:4:region:rnooverlap}. Moreover, we can make the angle between $e_1$ and $e_{2}$ in the interior cell acute in such a way that we fulfill \ref{inv:4:region:ep}. Once more, this adjustment may make a slight repositioning of $t_\mu$ necessary as argued above. We proceed symmetrically if $e_{k-1}$ has the property.  Second, it may occur that $k=2$, i.e, $\nu_1=\nu_{k-1}$. In this scenario, we necessarily have that $\mathcal{E}(\mu)$ is trapezoid-shaped. As a result, we reposition both $s_\mu$ and $t_\mu$ inside $\mathcal{E}_\mu$ as shown in~\cref{fig:sp-4-s-node-app-3}. Then, $(s_\mu,t_\mu)$ fulfills \ref{inv:4:region:r}, \ref{inv:4:region:rnooverlap} and \ref{inv:4:region:ep} for $\nu_1$.

In either case, we do not create crossings and we maintain all vertices on the outer cell which yields \ref{inv:4:rightangle} and~\ref{inv:4:outer}.  As we do not insert any other virtual edge corresponding to $S$-, we also satisfy \ref{inv:4:region:es1} and~\ref{inv:4:region:es2}.

\paragraph{Running time.} It remains to analyze the running time of the algorithm. As in \cref{thm:sp-degree-3}, we construct the SPQR-tree $\mathcal{T}$ in linear time due to~\cite{DBLP:conf/gd/GutwengerM00} and perform a constant number of operations per node, resulting in an overall linear running time.
\qed\end{proof}
\end{document}